\documentstyle[11pt,twoside]{article}

\begin{document}

\newcommand{\be}{\begin{equation}}
\newcommand{\ee}{\end{equation}}
\newcommand{\bea}{\begin{eqnarray}}
\newcommand{\eea}{\end{eqnarray}}
\newcommand{\beas}{\begin{eqnarray*}}
\newcommand{\eeas}{\end{eqnarray*}}
\newcommand{\bdm}{\begin{displaymath}}
\newcommand{\edm}{\end{displaymath}}
\newcommand{\ba}{\begin{array}}
\newcommand{\ea}{\end{array}}
\newcommand{\bi}{\begin{itemize}}
\newcommand{\ei}{\end{itemize}}
\newcommand{\ben}{\begin{enumerate}}
\newcommand{\een}{\end{enumerate}}
\newcommand{\bc}{\begin{center}}
\newcommand{\ec}{\end{center}}
\newcommand{\bfl}{\begin{flushleft}}
\newcommand{\efl}{\end{flushleft}}
\newcommand{\bfr}{\begin{flushright}}
\newcommand{\efr}{\end{flushright}}
\newcommand{\bd}{\begin{description}}
\newcommand{\ed}{\end{description}}
\newcommand{\bq}{\begin{quote}}
\newcommand{\eq}{\end{quote}}
\newcommand{\bfg}{\begin{figure}}
\newcommand{\efg}{\end{figure}}
\newcommand{\bt}{\begin{table}}
\newcommand{\et}{\end{table}}
\newcommand{\btb}{\begin{tabular}}
\newcommand{\etb}{\end{tabular}}
\newcommand{\btg}{\begin{tabbing}}
\newcommand{\etg}{\end{tabbing}}

\newcommand{\LLLLLA}{\Huge}
\newcommand{\LLLLA}{\huge}
\newcommand{\LLLA}{\LARGE}
\newcommand{\LLA}{\Large}
\newcommand{\LA}{\large}
\newcommand{\NS}{\normalsize}
\newcommand{\SM}{\small}
\newcommand{\FS}{\footnotesize}
\newcommand{\SS}{\scriptsize}
\newcommand{\T}{\tiny}

\newcommand{\Ds}{\displaystyle}
\newcommand{\Ts}{\textstyle}
\newcommand{\Fs}[1]{\mbox{\FS $#1$}}
\newcommand{\Ss}{\scriptstyle}
\newcommand{\SSs}{\scriptscriptstyle}

\newcommand{\e}{\enspace}
\newcommand{\bs}{\mbox{$\backslash$}}
\newcommand{\bm}[1]{\mbox{\boldmath $#1$}}
\newcommand{\itg}{\int \limits}
\newcommand{\oitg}{\oint \limits}
\newcommand{\eps}{\varepsilon}
\renewcommand{\Re}{\mbox{Re} \,}
\renewcommand{\Im}{\mbox{Im} \,}
\newcommand{\sign}{\mbox{sign} \,}
\newcommand{\Res}{\mbox{Res} \,}
\newcommand{\Li}{\mbox{Li}_{2}}
\newcommand{\f}[1]{\, \mbox{\raisebox{0.15ex}{\FS $#1$}} \,}
\newcommand{\newpages}{$\e$ \newpage \thispagestyle{empty} $\e$}

\newcommand{\N}{\mbox{\sf I \hspace{-1.7ex} N}}
\newcommand{\Z}{\mbox{\sf Z \hspace{-2.4ex} Z}}
\newcommand{\Q}{\mbox{\sf Q \hspace{-2.8ex} \raisebox{0.15ex}{\FS I}
                \hspace{0.1ex}}}
\newcommand{\R}{\mbox{\sf I \hspace{-1.7ex} R}}
\newcommand{\C}{\mbox{\sf C \hspace{-2.5ex} \raisebox{0.15ex}{\FS I}
                \hspace{-0.2ex}}}

\newcommand{\partialslash}
           {\mbox{$ \partial \hspace{-1.2ex} \mbox{/} \hspace{-0.05ex} $}}
\newcommand{\Aslash}
           {\mbox{$ A \hspace{-1.25ex} \mbox{/} \hspace{0ex} $}}
\newcommand{\Bslash}
           {\mbox{$ B \hspace{-1.45ex} \mbox{/} \hspace{0ex} $}}
\newcommand{\Dslash}
           {\mbox{$ D \hspace{-1.5ex} \mbox{/} \hspace{0ex} $}}
\newcommand{\Dcalslash}
           {\mbox{$ {\cal D} \hspace{-1.5ex} \mbox{/} \hspace{0ex} $}}
\newcommand{\Wslash}
           {\mbox{$ W \hspace{-1.9ex} \mbox{/} \hspace{0ex} $}}
\newcommand{\Xslash}
           {\mbox{$ X \hspace{-1.55ex} \mbox{/} \hspace{0ex} $}}
\newcommand{\Zslash}
           {\mbox{$ Z \hspace{-1.45ex} \mbox{/} \hspace{0ex} $}}
\newcommand{\aslash}
           {\mbox{$ a \hspace{-1.2ex} \mbox{/} \hspace{0ex} $}}
\newcommand{\bslash}
           {\mbox{$ b \hspace{-1ex} \mbox{/} \hspace{-0.12ex} $}}
\newcommand{\cslash}
           {\mbox{$ c \hspace{-0.9ex} \mbox{/} \hspace{-0.13ex} $}}
\newcommand{\dslash}
           {\mbox{$ d \hspace{-1.1ex} \mbox{/} \hspace{-0.05ex} $}}
\newcommand{\eslash}
           {\mbox{$ e \hspace{-1ex} \mbox{/} \hspace{-0.12ex} $}}
\newcommand{\kslash}
           {\mbox{$ k \hspace{-1.1ex} \mbox{/} \hspace{-0.07ex} $}}
\newcommand{\lslash}
           {\mbox{$ l \hspace{-0.9ex} \mbox{/} \hspace{-0.15ex} $}}
\newcommand{\pslash}
           {\mbox{$ p \hspace{-1ex} \mbox{/} \hspace{-0.08ex} $}}
\newcommand{\qslash}
           {\mbox{$ q \hspace{-1.1ex} \mbox{/} \hspace{-0.05ex} $}}
\newcommand{\vslash}
           {\mbox{$ v \hspace{-1.1ex} \mbox{/} \hspace{-0.07ex} $}}
\newcommand{\epsslash}
           {\mbox{$ \eps \hspace{-1.1ex} \mbox{/} \hspace{-0.05ex} $}}

\newcommand{\notRightarrow}
           {\mbox{$ \,\, \Rightarrow \hspace{-2.5ex} \mbox{/} \,\,\, $}}
\newcommand{\notLeftarrow}
           {\mbox{$ \,\, \Leftarrow \hspace{-2.1ex} \mbox{/} \,\,\, $}}
\newcommand{\notLeftrightarrow}
           {\mbox{$ \,\, \Leftrightarrow \hspace{-2.35ex} \mbox{/} \,\,\, $}}
\newcommand{\notLongrightarrow}
           {\mbox{$ \,\, \Longrightarrow \hspace{-3.3ex} \mbox{/}
            \,\,\,\,\, $}}
\newcommand{\notLongleftarrow}
           {\mbox{$ \,\, \Longleftarrow \hspace{-2.8ex} \mbox{/}
            \,\,\,\,\, $}}
\newcommand{\notLongleftrightarrow}
           {\mbox{$ \,\, \Longleftrightarrow \hspace{-3.3ex} \mbox{/}
            \,\,\,\,\, $}}

\title{\bfr \NS MZ-TH 93-11 \\[1cm] \efr
\bf Loop Integrals, ${\cal R}$ Functions \\
and their Analytic Continuation \\[2cm]}
\author{L. Br\"ucher${}^{1}$,
J. Franzkowski${}^{2}$,
D. Kreimer${}^{3}$ \\[1cm]
{\NS Institut f\"ur Physik} \\
{\NS Johannes Gutenberg-Universit\"at} \\
{\NS Staudingerweg 7} \\
{\NS D-55099 Mainz} \\
{\NS Germany} \\[2cm]}
\date{July 1993 \\[2cm]}

\maketitle
\begin{abstract}


$\e$ \\
{\NS To entirely determine the resulting functions of one-loop integrals
it is necessary to find the correct analytic continuation to
all relevant kinematical regions. We argue that this continuation
procedure may be performed in a general and mathematical accurate way
by using the ${\cal R}$ function notation of these integrals. The two-
and three-point cases are discussed explicitly in this manner.}

\end{abstract}

\footnotetext[1]{e-mail: Bruecher@vipmza.physik.uni-mainz.de}
\footnotetext[2]{e-mail: Franzkowski@vipmza.physik.uni-mainz.de}
\footnotetext[3]{e-mail: Kreimer@vipmza.physik.uni-mainz.de}

\thispagestyle{empty}

\newpage



\section{Introduction}

As pointed out recently (cf. \cite{Kr1}, \cite{Fra}, \cite{Kr2},
\cite{Kr3}) the integrals which may appear in one-loop Feynman diagrams
of any renormalizable field theory are expressible in terms of ${\cal
R}$ functions. This result is true despite the differing properties of
the various N-point functions and the numerous possibilities of placing
tensor structure in the numerator. We use \cite{Car} as a reference
for the class of ${\cal R}$ functions.


Making use of this feature the evaluation of one-loop N-point functions
is simplified drastically by virtue of the systematic recurrence
relations for ${\cal R}$ functions, which reduce the different tensor
type ${\cal R}$ functions to a set of fundamental ${\cal R}$ functions.
Each N-point function may be represented by such a characteristic
fundamental set. In this paper we restrict ourselves to two- and
three-point functions.

As it is well-known the results for loop integrals have to be discussed
in all kinematical regions of interest and often involve an analytic
continuation to all kinematical regimes at the end of the calculation.
Of course this statement is also true if one-loop functions are
expressed in terms of ${\cal R}$ functions. The analytic continuation
of ${\cal R}$ functions is performed in a very general way in
\cite{Car} using the analytical properties of this functions.
Unfortunately, the general formulary in \cite{Car} for the analytic
continuation procedure is not designed to our practical purposes,
because the  ${\cal R}$ functions of our interest -- though by
definition maximally analytical continued -- are expanded in one of
their parameters. They may be rewritten in terms of logarithms and
dilogarithms. In this paper we therefore develop the analytic
continuation formulae which are sufficient for the calculation of all
possible one-loop integrals. We hope to clarify some subtleties
involved in this procedure (cf. \cite{Sch}).

\section{Characteristics of ${\cal R}$ functions}

In the two-point case all integrals are expressed in linear combinations
of the two ${\cal R}$ functions (cf. \cite{Kr1}, \cite{Kr2})
\bea
\label{r1}
& {\cal R}_{1-\eps} ({\Ts - \frac{1}{2} + \eps}, 1; z_{1}, z_{2}) & \\
\label{r2}
& {\cal R}_{-\eps} ({\Ts - \frac{1}{2} + \eps}, 1; z_{1}, z_{2}) &
\eea
where the second function (\ref{r2}) is the solution of the scalar
integral and the first one only appears in higher tensor cases. In our
notation $\eps = \frac{(4 - D)}{2}$ represents the usual dimensional
regularization parameter and the $z_{i}$ contain the parameters of the
integral -- internal masses and external momentum components.

The three-point functions are described by three ${\cal R}$ functions:
\bea
\label{r3}
& {\cal R}_{1-2\eps}(\eps, \eps, 1; z_{1}, z_{2}, z_{3}) & \\
\label{r4}
& {\cal R}_{-2\eps}(\eps, \eps, 1; z_{1}, z_{2}, z_{3}) & \\
\label{r5}
& {\cal R}_{-1-2\eps}(\eps, \eps, 1; z_{1}, z_{2}, z_{3}) &
\eea
Here again the second function (\ref{r4}) is the scalar solution,
whereas the others appear only in the case of tensor integrals.

According to \cite{Car} ${\cal R}_{t}(b;z) = {\cal R}_{t}(b_{1},
\ldots,b_{k};z_{1},\ldots,z_{k})$ is defined for arbitrary complex
index $t$, whereas the parameters $b = (b_{1}, \ldots, b_{k})$ have to
satisfy the restriction
\be
\label{restr}
\sum_{i=1}^{k} b_{i} \ne 0, -1, -2, \ldots
\ee
The cut due to the arguments $z = (z_{1}, \ldots, z_{k})$ is usually
chosen to lie on the negative real axis, which means that the function
is defined uniquely as far as $|\arg z_{i}| < \pi \e \forall i$. The
most important transformation laws for ${\cal R}$ functions are
summarized in appendix \ref{Form} of this paper.

In the limit $\eps \to 0$ the restriction of parameters (\ref{restr})
is fulfilled in every function (\ref{r1}) -- (\ref{r5}) and the $z_{i}$
do not touch the cut provided the $i \eta$ prescription in the
propagators was not neglected during calculations, whereas any
transformation of arguments -- especially concerning the scaling law
(\ref{skal}) -- has to be performed carefully not to cross the cut for
any of the $z_{i}$.

Up to now there was no need to require any kinematical restriction --
except the quite natural assumption that masses and momentum components
are real valued parameters of the integrals. In this context we claim
that the ${\cal R}$ functions (\ref{r1}) -- (\ref{r5}) carry the entire
information in all kinematical regions.

\section{The two-point case}

To evaluate the two-point integrals for concrete values it is necessary
to expand the ${\cal R}$ functions (\ref{r1}), (\ref{r2}) in powers of
$\eps$, starting with the scalar ${\cal R}$ function (\ref{r2}) and
bearing in mind the two different cases
\bea
\label{ra}
& {\cal R}_{-\eps} ({\Ts - \frac{1}{2} + \eps},1;-x_{1}+i\rho_{1},
-x_{2}+i\rho_{2})
& \\
\label{rb}
& {\cal R}_{-\eps} ({\Ts - \frac{1}{2} + \eps},1;-x_{1}+i\rho_{1},
-x_{2}-i\rho_{2}) &
\eea
whereas the $x_{i}$ and $\rho_{i}$ are supposed to be real and positive
and the $\rho_{i}$ are understood to be infinitesimal. The given
arguments are the only possibilities which might appear in two-point
functions (cf. \cite{Kr2}). The usual procedure, following \cite{Kr2},
makes use now first of the scaling law (\ref{skal}) and then of Euler's
transformation (\ref{euler}). Of course the scaling law is save for
(\ref{ra}), because both arguments of the ${\cal R}$ function lie in
the same (the upper) half plane and none of them is forced to cross the
cut during the scaling process. The whole transformation procedure of
\cite{Kr2} may therefore be applied without any complications.
(\ref{skal}) and (\ref{euler}) change the function to
\bea
\lefteqn{{\cal R}_{-\eps} ({\Ts - \frac{1}{2} +
\eps},1;-x_{1}+i\rho_{1},-x_{2}+i\rho_{2})} \nonumber \\
& = & (-x_{1}+i\rho_{1})^{1-\eps} \, (-x_{2}+i\rho_{2})^{-1} \, {\cal
R}_{-\frac{1}{2}} \left(- \frac{1}{2} + \eps,1;1,\frac{x_{1}-i\rho_{1}}{x_{2}
-i\rho_{2}}\right)
\eea
The arguments of the ${\cal R}$ function lie in the right half plane so
that the assumptions of the first quadratic transformation are
fulfilled and the evaluation is resulting in
\bea
\label{rc}
\lefteqn{{\cal R}_{-\eps} ({\Ts - \frac{1}{2} + \eps},1;-x_{1}+i\rho_{1},
-x_{2}+i\rho_{2}) = {\Ts \frac{1}{2}} \, (-x_{1}+i\rho_{1})^{-\eps} \,
\left({\Ts \frac{-x_{1}+i\rho_{1}}{-x_{2}+i\rho_{2}}}\right)^{1-\eps}} \\
& & \times \left[\left(1+\sqrt{1-{\Ts
\frac{-x_{1}+i\rho_{1}}{-x_{2}+i\rho_{2}}}}\right)^{-1+2\eps} + \left(1
-\sqrt{1-{\Ts \frac{-x_{1}+i\rho_{1}}{-x_{2}+i\rho_{2}}}}\right)^{-1+2\eps}
\right] + O(\eps^{2})
\nonumber
\eea

The situation is completely different for the second type of ${\cal R}$
function (\ref{rb}). To be able to apply the scaling law we first
perform a transformation using the integral representation (\ref{int}):
\bea
\lefteqn{{\cal R}_{-\eps} ({\Ts - \frac{1}{2} + \eps},1;-x_{1}+i\rho_{1},-x_{2}
-i\rho_{2})} \\
& = & \frac{1}{B(\eps,\frac{1}{2})} \, \itg_{0}^{\infty} \!
\frac{s^{-\frac{1}{2}} \, ds}{(s-x_{1}+i\rho_{1})^{-\frac{1}{2}+\eps} (s-x_{2}
-i\rho_{2})} \nonumber
\eea
We now change the integration contour in the following manner (from (a)
to (b)),
\unitlength=1.00mm
\special{em:linewidth 0.4pt}
\linethickness{0.4pt}
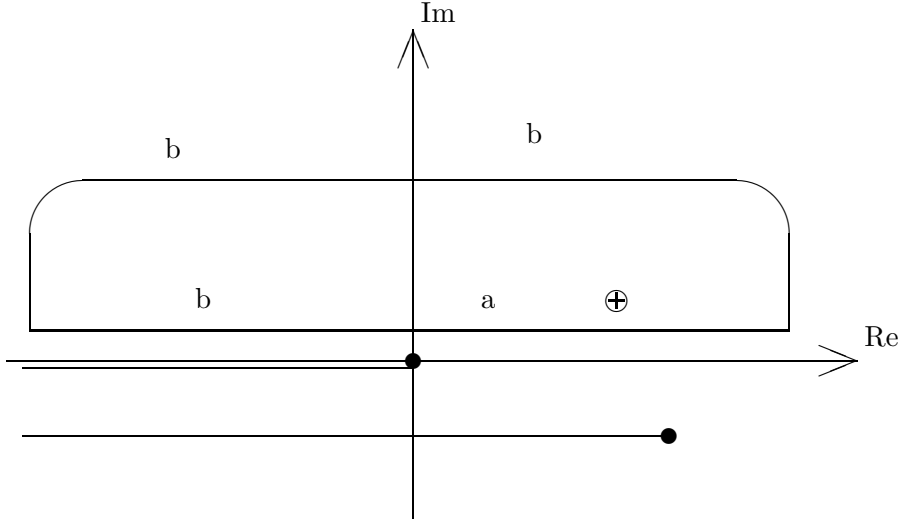
\begin{picture}(121.00,76.00)
\put(60.50,30.00){\oval(101.00,40.00)[t]}
\put(7.00,26.00){\line(1,0){108.00}}
\put(115.00,26.00){\line(1,0){5.00}}
\put(120.00,26.00){\line(-5,2){5.00}}
\put(61.00,5.00){\line(0,1){65.00}}
\put(61.00,70.00){\line(2,-5){2.00}}
\put(61.00,26.00){\circle*{2.00}}
\put(61.00,25.00){\line(-1,0){52.00}}
\put(111.00,30.00){\line(-1,0){101.00}}
\put(95.00,16.00){\circle*{2.00}}
\put(95.00,16.00){\line(-1,0){86.00}}
\put(88.00,34.00){\circle{2.83}}
\put(121.00,28.00){\makebox(0,0)[lb]{Re}}
\put(62.00,71.00){\makebox(0,0)[lb]{Im}}
\put(70.00,33.00){\makebox(0,0)[lb]{a}}
\put(76.00,55.00){\makebox(0,0)[lb]{b}}
\put(28.00,53.00){\makebox(0,0)[lb]{b}}
\put(32.00,33.00){\makebox(0,0)[lb]{b}}
\put(120.00,26.00){\line(-5,-2){5.00}}
\put(88.00,35.00){\line(0,-1){2.00}}
\put(89.00,34.00){\line(-1,0){2.00}}
\put(61.00,70.00){\line(-2,-5){2.00}}
\end{picture}
$\e$ \\
applying the residue theorem. No cut is crossed during this procedure
and the contribution at infinity is regulated out by $\eps$, so that
the integral takes on the form
\bea
\lefteqn{{\cal R}_{-\eps} ({\Ts - \frac{1}{2} + \eps},1;-x_{1}+i\rho_{1},
-x_{2}-i\rho_{2})} \\
& \hspace*{-0.2cm} = & \hspace*{-0.3cm} \frac{1}{B(\eps,\frac{1}{2})} \,
\left[\frac{2 \pi i \, (x_{2}+i\rho_{2})^{-\frac{1}{2}}}{(x_{2}-x_{1}
+i\rho_{1} + i\rho_{2})^{-\frac{1}{2}+\eps}} - \itg_{-\infty}^{0} \!
\frac{(s+i\rho)^{-\frac{1}{2}} \, ds}{(s-x_{1}+i\rho_{1})^{-\frac{1}{2}+\eps}
(s-x_{2}-i\rho_{2})} \right]
\nonumber \\
& \hspace*{-0.2cm} = & \hspace*{-0.3cm} \frac{1}{B(\eps,\frac{1}{2})} \,
\left[\frac{2 \pi i \, (x_{2}+i\rho_{2})^{-\frac{1}{2}}}{(x_{2}-x_{1}
+i\rho_{1} + i\rho_{2})^{-\frac{1}{2}+\eps}} + \itg_{0}^{\infty} \!
\frac{(-s+i\rho)^{-\frac{1}{2}} \, ds}{(-s-x_{1}+i\rho_{1})^{-\frac{1}{2}
+\eps} (s+x_{2}+i\rho_{2})} \right]
\nonumber \\
& \hspace*{-0.2cm} = & \hspace*{-0.3cm} \frac{1}{B(\eps,\frac{1}{2})} \,
\left[\frac{2 \pi i \, (x_{2}+i\rho_{2})^{-\frac{1}{2}}}{(x_{2}-x_{1}
+i\rho_{1} + i\rho_{2})^{-\frac{1}{2}+\eps}} + e^{-i\pi\eps} \itg_{0}^{\infty}
\! \frac{s^{-\frac{1}{2}} \, ds}{(s+x_{1}-i\rho_{1})^{-\frac{1}{2}+\eps}
(s+x_{2}+i\rho_{2})} \right] \nonumber \\
& \hspace*{-0.2cm} = & \hspace*{-0.3cm} \frac{2 \pi i}{B(\eps,\frac{1}{2})} \,
\frac{(x_{2}+i\rho_{2})^{-\frac{1}{2}}}{(x_{2}-x_{1}+i\rho_{1} + i\rho_{2})
^{-\frac{1}{2}+\eps}} + e^{-i\pi\eps} \, {\cal R}_{-\eps} ({\Ts - \frac{1}{2}
+ \eps},1;x_{1}-i\rho_{1},x_{2}+i\rho_{2}) \nonumber
\eea
The resulting ${\cal R}$ function is save with respect to the scaling law
because both arguments lie near the positive real axis. So by virtue of
(\ref{skal}) and (\ref{euler}) this ${\cal R}$ function leads to (cf.
\cite{Kr2})
\bea
\lefteqn{{\cal R}_{-\eps} ({\Ts - \frac{1}{2} +
\eps},1;x_{1}-i\rho_{1},x_{2}+i\rho_{2})} \nonumber \\
& = & (x_{1}-i\rho_{1})^{1-\eps} \, (x_{2}+i\rho_{2})^{-1} \, {\cal R}
_{-\frac{1}{2}} \left(- \frac{1}{2} + \eps,1;1,\frac{x_{1}-i\rho_{1}}{x_{2}
+i\rho_{2}}\right)
\eea
The arguments lie in the right half plane so that now -- according to
\cite{Kr2} -- the first quadratic transformation (\ref{quad}) may be
applied and the whole function (\ref{rb}) is transformed to
\bea
\lefteqn{{\cal R}_{-\eps} ({\Ts - \frac{1}{2} + \eps},1;-x_{1}+i\rho_{1},
-x_{2}-i\rho_{2})} \nonumber \\
& = & \frac{2 \pi i}{B(\eps,\frac{1}{2})} \,
\frac{(x_{2}+i\rho_{2})^{-\frac{1}{2}}}{(x_{2}-x_{1}+i\rho_{1} + i\rho_{2})
^{-\frac{1}{2}+\eps}} + e^{-i\pi\eps} \, {\Ts \frac{1}{2}} \, (x_{1}
-i\rho_{1})^{-\eps} \, \left({\Ts \frac{x_{1}-i\rho_{1}}{x_{2}+i\rho_{2}}}
\right)^{1-\eps} \nonumber \\
& & \times \left[\left(1+\sqrt{1-{\Ts \frac{x_{1}-i\rho_{1}}{x_{2}
+i\rho_{2}}}}\right)^{-1+2\eps} + \left(1-\sqrt{1-{\Ts \frac{x_{1}
-i\rho_{1}}{x_{2}+i\rho_{2}}}}\right)^{-1+2\eps}\right] + O(\eps^{2})
\nonumber \\
\label{rd}
& = & \frac{2 \pi i}{B(\eps,\frac{1}{2})} \, \frac{(x_{2}+i\rho_{2})
^{-\frac{1}{2}}}{(x_{2}-x_{1}+i\rho_{1} + i\rho_{2})^{-\frac{1}{2}}} +
{\Ts \frac{1}{2}} \, (-x_{1}+i\rho_{1})^{-\eps} \, \left({\Ts
\frac{x_{1}-i\rho_{1}}{x_{2}+i\rho_{2}}}\right)^{1-\eps} \\
& & \times \left[\left(1+\sqrt{1-{\Ts
\frac{x_{1}-i\rho_{1}}{x_{2}+i\rho_{2}}}}\right)^{-1+2\eps} + \left(1
-\sqrt{1-{\Ts \frac{x_{1}-i\rho_{1}}{x_{2}+i\rho_{2}}}}\right)^{-1+2\eps}
\right] + O(\eps^{2})
\nonumber
\eea
Putting together both cases (\ref{rc}) and (\ref{rd}) we get
\bea
\lefteqn{{\cal R}_{-\eps} ({\Ts - \frac{1}{2} + \eps}, 1; z_{1}, z_{2})
= \frac{2\pi i}{B(\eps,\frac{1}{2})} \, \frac{\sqrt{z_{1}-z_{2}}}{\sqrt{
-z_{2}}} \, \theta(\Im(-z_{2}))} \\
& & + {\Ts \frac{1}{2}} \, z_{1}^{-\eps} \, \left(\frac{z_{1}}{z_{2}}\right)
^{1-\eps} \, \left[\left(1+\sqrt{1-\frac{z_{1}}{z_{2}}}\right)^{-1+2\eps} +
\left(1-\sqrt{1-\frac{z_{1}}{z_{2}}}\right)^{-1+2\eps}\right] +
O(\eps^{2}) \nonumber
\eea

The other two-point ${\cal R}$ function (\ref{r1}) is evaluated faster
thanks to the already solved function (\ref{r2}), using the parameter
shifting formula (\ref{shift}):
\bea
\lefteqn{{\cal R}_{1-\eps} ({\Ts - \frac{1}{2} + \eps}, 1; z_{1},
z_{2})} \\
& = & 2 [({\Ts -\frac{1}{2}}+\eps) \, {\cal R}_{1-\eps} ({\Ts - \frac{1}{2}
+ \eps},0; z_{1}, z_{2}) + (1-\eps)y \, {\cal R}_{-\eps} ({\Ts - \frac{1}{2} +
\eps},1; z_{1}, z_{2})] \nonumber
\eea
The first ${\cal R}$ function on the right hand side is simply
$z_{1}^{1-\eps}$, the second one is the function (\ref{r2}) which we
solved before.

\section{The three-point case}

For the three-point integrals we employ an alternative strategy.
Beginning with the scalar function (\ref{r4}) we apply the series
expansion (\ref{entw})
\bea
\lefteqn{{\cal R}_{-2\eps}(\eps,\eps,1;x,y,z) \, = \,
\sum_{n=0}^{\infty} \, \frac{(2\eps,n)}{(2\eps + 1,n)} \sum_{m_{1}=0}^{n} \,
\sum_{m_{2}=0}^{n-m_{1}} \frac{(\eps,m_{1})}{m_{1}!} \, \frac{(\eps,m_{2})}
{m_{2}!}} \nonumber \\
& & \hspace{1.5cm} \frac{(1,n - m_{1} - m_{2})}{(n - m_{1} - m_{2})!} \,\,
(1 - x)^{m_{1}} \, (1 - y)^{m_{2}} \, (1 - z)^{n-m_{1}-m_{2}} \\
& = & \hspace{-0.2cm} 1 + 2\eps \, \sum_{n=1}^{\infty} \, \frac{1}{2\eps + n}
\sum_{m_{1}=0}^{n} \sum_{m_{2}=0}^{n-m_{1}} \frac{(\eps,m_{1})}{m_{1}!} \,
\frac{(\eps,m_{2})}{m_{2}!} \,\, (1 - x)^{m_{1}} \, (1 - y)^{m_{2}} \, (1 - z)
^{n-m_{1}-m_{2}} \nonumber
\eea
Appell's symbol $(\eps,m)$ we define as in (\ref{app}):
\bdm
(\eps,m) = \eps (\eps+1) \cdots (\eps+m-1)
\edm
It is of order $\eps$, except the case $m =
0$, where $(\eps,m) = 1$. Therefore we derive -- neglecting the
contributions of higher order than $O(\eps^{2})$:
\bea
\lefteqn{{\cal R}_{-2\eps}(\eps,\eps,1;x,y,z) = 1 + 2 \eps \, \sum_{n=1}
^{\infty} \, \frac{1}{n} \, \left(1 - \frac{2\eps}{n}\right) \,
(1 - z)^{n}} \nonumber \\
& & + 2 \eps^{2} \, \sum_{n=1}^{\infty} \, \frac{1}{n}
\sum_{m_{2}=1}^{n} \frac{(m_{2}-1)!}{m_{2}!} \, (1 - y)^{m_{2}} \, (1 - z)
^{n-m_{2}} \nonumber \\
& & + 2 \eps^{2} \, \sum_{n=1}^{\infty} \, \frac{1}{n}
\sum_{m_{1}=1}^{n} \frac{(m_{1}-1)!}{m_{1}!} \, (1 - x)^{m_{1}} \, (1 - z)
^{n-m_{1}} + O(\eps^{3}) \nonumber \\
& = & 1 - 2 \eps \ln z - 4 \eps^{2} \, \Li (1 - z) + 2 \eps^{2} \,
\sum_{n=1}^{\infty} \, \frac{1}{n} \sum_{m_{2}=1}^{n} \frac{1}{m_{2}} \,
(1 - y)^{m_{2}} \, (1 - z)^{n-m_{2}} \nonumber \\
\label{result}
& & + 2 \eps^{2} \, \sum_{n=1}^{\infty} \, \frac{1}{n}
\sum_{m_{1}=1}^{n} \frac{1}{m_{1}} \, (1 - x)^{m_{1}} \, (1 - z)
^{n-m_{1}} + O(\eps^{3}) \\
& = & z^{-2\eps} \, \left\{1 + 2\eps^{2} \, \left[\Li \left(1 -
\frac{x}{z}\right) + \Li \left(1 - \frac{y}{z}\right) \right. \right.
\nonumber \\
& & + \ln \left(1 - \frac{x}{z}\right) \eta \left(x, \frac{1}{z}\right)
+ \ln \left(1 - \frac{y}{z}\right) \eta \left(y, \frac{1}{z}\right) +
\ln z \, \left[\eta\left(x - z,\frac{1}{1-z}\right) \right. \nonumber \\
& & \left. \left. \left. - \eta\left(x - z,-\frac{1}{z}\right) + \eta\left(y
- z,\frac{1}{1-z}\right) - \eta\left(y - z,-\frac{1}{z}\right)\right]\right]
\right\} + O(\eps^{3}) \nonumber
\eea
In the last transformation we used the sum formula
\bea
\lefteqn{\sum_{n=1}^{\infty} \, \frac{1}{n} \, \sum_{m=1}^{n} \, \frac{1}{m} \,
(1 - x)^{m} \, (1 - y)^{n-m} = \Li (1 - y) + \Li \left(1 - \frac{x}{y}
\right) + \frac{1}{2} \, (\ln y)^{2}} \nonumber \\
\label{sum}
& & + \ln y \, \left[\eta\left(x - y,\frac{1}{1 - y}\right)
- \eta\left(x - y, - \frac{1}{y}\right) \right] + \ln \left(1
- \frac{x}{y}\right) \eta\left(x,\frac{1}{y}\right)
\eea
which we prove in appendix \ref{Proof}. In the notation of the
dilogarithm function $\Li (z)$ we follow the convention of the standard
reference \cite{Lew}. The $\eta$ function is the usual abbreviation for
\bea
\eta(a,b) & = & 2\pi i \left[\theta(-\Im a) \theta(-\Im b) \theta(\Im (a b))
\right. \nonumber \\
& & \hspace{0.35cm} \left. - \theta(\Im a) \theta(\Im b) \theta(-\Im (a b))
\right]
\eea

It should be emphasized that in spite of the small convergence domain
of the series (\ref{entw}) the result is valid in the whole complex
plane -- except the cuts of logarithms and dilogarithms -- by help of
the uniqueness of the analytic continuation procedure.

If the calculation is done using the scaling law (\ref{skal}), writing
\be
{\cal R}_{-2\eps}(\eps,\eps,1;x,y,z) = z^{-2\eps} \, {\cal R}_{-2\eps}
(\eps,\eps,1; {\Ts \frac{x}{z},\frac{y}{z}},1)
\ee
the whole evaluation simplifies -- using again the expansion
(\ref{entw}) -- but the result is restricted to a smaller domain:
\be
{\cal R}_{-2\eps}(\eps,\eps,1;x,y,z) = z^{-2\eps} \, \left\{1 + 2\eps^{2} \,
\left[\Li \left(1 - \frac{x}{z}\right) + \Li \left(1 - \frac{y}{z}\right)
\right] \right\} + O(\eps^{3})
\ee
This formula is identical to the former one (\ref{result}) -- as long
as every $\eta$ function vanishes -- or from another point of view: the
$\eta$ terms in (\ref{result}) represent the corrections of the scaling
law if one argument of the ${\cal R}$ function crosses the cut during
the transformation of arguments $x \to \frac{x}{z}, \, y \to
\frac{y}{z}$ respectively.

The entire scalar three-point function (cf. \cite{Kr1}, \cite{Kr3})
consists of six ${\cal R}$ functions of the form (\ref{r4}) and
some prefactors. It therefore involves twelve dilogarithms and a considerable
amount of $\eta$ functions if the result (\ref{result}) is substituted
-- the same number of dilogarithms as the standard publication of 't
Hooft and Veltman \cite{Hoo} does. The equivalence of both results was
checked in all different kinematical regions numerically.

The first of the three-point ${\cal R}$ functions (\ref{r3}) may be
evaluated in the following way, making use of the already expanded
second one:
\be
{\cal R}_{1-2\eps}(\eps, \eps, 1; x, y, z) = 2 \eps \, {\cal R}_{1-2\eps}
(\eps, \eps, 0; x, y, z) + (1 - 2 \eps) z \, {\cal R}_{-2\eps}(\eps, \eps,
1; x, y, z)
\ee
due to the parameter shifting formula (\ref{shift}). The two ${\cal R}$
functions on the right hand side are easier to handle: the first one
has the advantage that it needs to be expanded only up to $O(\eps)$ -- by
virtue of (\ref{entw}) -- and the second one was already expanded in
(\ref{result}). The complete result is
\bea
\lefteqn{{\cal R}_{1-2\eps}(\eps, \eps, 1; x, y, z) = (1 - 2 \eps) z^{1-2\eps}
+ \eps (x + y)} \nonumber \\
& & + 2\eps^{2} \left[- y \, \ln y - x \, \ln x + z \, \Li\left(1 -
\frac{x}{z}\right) + z \, \Li\left(1 - \frac{y}{z}\right)\right. \\
& & + z \, \ln \left(1 - \frac{x}{z}\right) \eta \left(x, \frac{1}{z}\right)
+ z \, \ln \left(1 - \frac{y}{z}\right) \eta \left(y, \frac{1}{z}\right) +
z \, \ln z \, \left[\eta\left(x - z,\frac{1}{1-z}\right) \right. \nonumber \\
& & \left. \left. - \eta\left(x - z,-\frac{1}{z}\right) + \eta\left(y
- z,\frac{1}{1-z}\right) - \eta\left(y - z,-\frac{1}{z}\right)\right]\right]
+ O(\eps^{3}) \nonumber
\eea

The third function (\ref{r5}) may be evaluated trivially applying
(\ref{euler}) and recognizing ${\cal R}_{0}(b;z) = 1$ in any case:
\be
{\cal R}_{-1-2\eps}(\eps, \eps, 1; z_{1}, z_{2}, z_{3}) = z_{1}^{-\eps}
\, z_{2}^{-\eps} \, z_{3}^{-1}
\ee

\section{R\'esum\'e and outlook}

We want to stress once more that, taking into account the correct
analytic continuation of ${\cal R}$ functions, our results agree with
the well known standard results for the one-loop two- and three-point
functions (cf. \cite{Hoo}, \cite{Pas}) in all kinematical regions.

We can also report the fact, that we were able to express arbitrary
four-point functions in terms of ${\cal R}$ functions (cf. \cite{Fra}).
The structure of the involved ${\cal R}$ functions coincides with the
discussed functions (\ref{r3}) -- (\ref{r5}) of the three-point case.
This four-point procedure may be also extended to five- and higher
N-point functions without complications. The results will be published
later.

\section*{Acknowledgements}

We thank J. G. K\"orner and K. Schilcher for many discussions on the
subject of this paper.

We thank U. Nierste for helpful discussions on the analytic structure of
four- and higher N-point functions.

\begin{appendix}

\setcounter{equation}{0}
\renewcommand{\theequation}{\mbox{A}.\arabic{equation}}


\newpage

$\e$ \\
{\LLA \bf Appendix}

\section{${\cal R}$ function formulary} \label{Form}

Using the following abbreviations
\bea
\beta & = & \sum_{i=1}^{k} \, b_{i} \\
\label{app}
(a,n) & = & \frac{\Gamma(a + n)}{\Gamma(a)} \\
b \pm e_{i} & = & (b_{1},\ldots,b_{i} \pm 1,\ldots,b_{k})
\eea
we review the most important transformation rules for ${\cal R}$
functions according to \cite{Car}:
\bi
\item (analytic continued) integral representation:
      \bea
      \lefteqn{\itg_{r}^{\infty} \! (x - r)^{\alpha - 1} \, \prod_{i=1}^{k} \,
      (z_{i} + w_{i} x)^{-b_{i}} \, dx} \nonumber \\
      \label{int}
      & = & B(\beta - \alpha, \alpha) \, {\cal R}_{\alpha - \beta}
      \left(b_{1},\ldots,b_{k};r + \frac{z_{1}}{w_{1}},\ldots, r +
      \frac{z_{k}}{w_{k}}\right) \, \prod_{i=1}^{k} \, w_{i}^{-b_{i}}
      \eea
\item scaling law:
      \be
      \label{skal}
      {\cal R}_{t}(b_{1},\ldots,b_{k};\lambda z_{1},\ldots,\lambda z_{k}) =
      \lambda^{t} {\cal R}_{t}(b_{1},\ldots,b_{k};z_{1},\ldots,z_{k})
      \ee
\item shifting parameters:
      \be
      \label{shift}
      \beta {\cal R}_{t}(b;z) = (\beta + t) {\cal R}_{t}(b + e_{i};z) - t
      z_{i}{\cal R}_{t-1}(b + e_{i};z); \qquad i \f{\in} (1, \ldots, k)
      \ee
\item series expansion:
      \bea
      \label{entw}
      \frac{{\cal R}_{t}(b;z)}{\Gamma(\beta)} & = & \frac{1}{\Gamma(\beta)} \,
      \sum_{n=0}^{\infty} \, \frac{(-t,n)}{(\beta,n)} \sum_{\{m_{i}\}}
      \frac{(b_{1},m_{1})}{m_{1}!} \cdots \frac{(b_{k},m_{k})}{m_{k}!} \\
      & & \hspace{1.6cm} \times (1 - z_{1})^{m_{1}} \cdots (1 - z_{k})^{m_{k}}
      \nonumber
      \eea
      keeping in mind that the right hand side converges only if $|1 - z_{i}|
      < 1 \e \forall i$. The sum $\{m_{i}\}$ is meant to be taken over all
      $m_{i}$ separately with the only restriction that $\sum_{i=1}^{k}
      m_{i} = n$
\item Euler's transformation:
      \be
      \label{euler}
      {\cal R}_{t}(b_{1},\ldots,b_{k};z_{1},\ldots,z_{k}) = {\cal R}_{-t-\beta}
      \left(b_{1},\ldots,b_{k};\frac{1}{z_{1}},\ldots,\frac{1}
      {z_{k}}\right) \, \prod_{i=1}^{k} z_{i}^{-b_{i}}
      \ee
\item first quadratic transformation:
      \be
      \label{quad}
      {\cal R}_{2t}(b,b;x,y) = {\cal R}_{t}\left(b+t,\frac{1}{2}-t;\left(
      \frac{x+y}{2}\right)^{2},xy\right)
      \ee
      valid only for $\Re x, \, \Re y > 0$ and $b + \frac{1}{2} \ne
      0,-1,-2,\ldots$.
\ei

\section{Proof of formula (\protect\ref{sum})} \label{Proof}

Rewriting the sums of the left hand side of (\ref{sum}) in the form
$\sum_{m=1}^{n} = \sum_{m=1}^{\infty} - \sum_{m=n+1}^{\infty}$ we obtain
\bea
\lefteqn{\sum_{n=1}^{\infty} \, \frac{1}{n} \, \sum_{m=1}^{n} \, \frac{1}{m} \,
(1 - x)^{m} \, (1 - y)^{n-m} = \sum_{n=1}^{\infty} \, \frac{1}{n} \,
(1 - y)^{n} \, \sum_{m=1}^{\infty} \, \frac{1}{m} \, \left(\frac{1 - x}
{1 - y}\right)^{m}} \\
& \hspace{4cm} & - \sum_{n=1}^{\infty} \, \frac{1}{n} \, (1 - y)^{n} \,
\sum_{m=0}^{\infty} \, \frac{1}{m+n+1} \, \left(\frac{1 - x}{1 -
y}\right)^{m+n+1} \nonumber
\eea
Applying formula ~5.2.3.4 of \cite{Pru}:
\be
\sum_{k=0}^{\infty} \, \frac{x^{k+a}}{k+a} = \itg_{0}^{x} \!
\frac{t^{a-1}}{1-t} \, dt \qquad (\Re a > 0)
\ee
we obtain
\bea
\lefteqn{\sum_{n=1}^{\infty} \, \frac{1}{n} \, \sum_{m=1}^{n} \, \frac{1}{m} \,
(1 - x)^{m} \, (1 - y)^{n-m}} \nonumber \\
& = & \ln y \, \ln\left(1 - \frac{1 - x}{1 - y}\right) -
\itg_{0}^{\Ts \frac{1-x}{1-y}} \!\! \frac{dt}{1 - t} \,
\sum_{n=1}^{\infty} \frac{1}{n} \, [(1 - y) t]^{n} \\
& = & \ln y \, \ln\left(\frac{x - y}{1 - y} \right) +
\itg_{0}^{\Ts \frac{1-x}{1-y}} \!\! \frac{dt}{1 - t} \, \ln(1 - (1 - y) t)
\nonumber \\
& = & \ln y \, \ln\left(\frac{x - y}{1 - y}\right) - \ln x \, \ln\left(\frac{y
- x}{y}\right) - \Li\left(\frac{x}{y}\right) + \Li\left(\frac{1}{y}\right)
\nonumber
\eea
Applying now two well-known transformation rules for dilogarithms (cf.
\cite{Hoo}):
\bea
\Li (z) & = & - \Li (1 - z) - \ln (z) \, \ln (1 - z) + \frac{\pi^{2}}{6} \\
\Li (z) & = & - \Li \left(\frac{1}{z}\right) - \frac{1}{2} \,
\left[\ln (- z)\right]^{2} - \frac{\pi^{2}}{6}
\eea
allows us to transform the dilogarithms into the form of (\ref{sum}).

\end{appendix}


\end{document}